\documentclass[twocolumn]{emulateapj}
\submitted{Science with Parkes @ 50 Years Young, 31 Oct. -- 4 Nov., 2011}

\shorttitle{The Methanol Multibeam Survey}
\shortauthors{Green et al.}

\begin{document}

\title{The Methanol Multibeam Survey}

\author{J. A. Green}
\affil{CSIRO Astronomy and Space Science, Australia Telescope National Facility,\\ PO Box 76, Epping, NSW 2121, Australia}
\email{james.green@csiro.au}

\author{and the Methanol Multibeam Survey Collaboration}
\affil{J. L. Caswell\altaffilmark{1}, G. A. Fuller\altaffilmark{2}, A. Avison\altaffilmark{2}, S. L. Breen\altaffilmark{1,3}, K. Brooks\altaffilmark{1},\\ 
      M. G. Burton\altaffilmark{4}, A. Chrysostomou\altaffilmark{5}, J. Cox\altaffilmark{6}, P. J. Diamond\altaffilmark{1,2},\\ 
      S. P. Ellingsen\altaffilmark{3}, M. D. Gray\altaffilmark{2}, M. G. Hoare\altaffilmark{7}, M. R. W. Masheder\altaffilmark{8}, \\ N. M. McClure-Griffiths\altaffilmark{1}, 
      M. Pestalozzi\altaffilmark{9}, C. Phillips\altaffilmark{1}, L. Quinn\altaffilmark{2},\\ M. A. Thompson\altaffilmark{5}, 
      M. A. Voronkov\altaffilmark{1}, A. Walsh$^{9}$, D. Ward-Thompson\altaffilmark{6}, \\
      D. Wong-McSweeney\altaffilmark{2}, J. A. Yates\altaffilmark{11} 
      and R. J. Cohen\altaffilmark{2}}
      
\altaffiltext{1}{same institute as J. A. Green}
\altaffiltext{2}{Jodrell Bank Centre for Astrophysics, Alan Turing Building, University of Manchester, Manchester, M13 9PL, UK}
\altaffiltext{3}{School of Mathematics and Physics, University of Tasmania, PrivateBag 37, Hobart, TAS 7001, Australia}
\altaffiltext{4}{School of Physics, University of New South Wales, Sydney, NSW 2052, Australia}
\altaffiltext{5}{Centre for Astrophysics Research, Science and Technology Research Institute, University of Hertfordshire, College Lane, Hatfield, AL10 9AB, UK}
\altaffiltext{6}{Department of Physics and Astronomy, Cardiff University, 5 The Parade, Cardiff, CF24 3YB, UK}
\altaffiltext{7}{School of Physics and Astronomy, University of Leeds, Leeds, LS2 9JT, UK}
\altaffiltext{8}{Astrophysics Group, Department of Physics, Bristol University, Tyndall Avenue, Bristol, BS8 1TL, UK}
\altaffiltext{9}{INAF -- Istituto Fisica Spazio Interplanetario, via del Fosso del Cavaliere 100, I-00133 Roma, Italy}
\altaffiltext{10}{School of Maths, Physics and IT, James Cook University, Townsville, QLD 4811, Australia}
\altaffiltext{11}{University College London, Department of Physics and Astronomy, Gower Street, London, WC1E 6BT, UK}











\begin{abstract}
A purpose built 7-beam methanol receiver, installed on the Parkes Radio Telescope, was used to survey the Galactic plane for newly forming high mass stars, pinpointed by strong methanol maser emission at 6.7 GHz. The Methanol Multibeam (MMB) survey observed over 60\% of the Galactic plane, detecting close to 1000 sources. The MMB survey provides a huge resource for studies of high-mass star formation, an important stage in the evolution of the interstellar medium. The MMB survey is also a valuable resource for investigations into the structure and dynamics of our Galaxy: with narrow velocity ranges of emission (typically only a few km/s) and velocities closely correlated with the systemic velocity of their surrounding molecular clouds, 6.7-GHz methanol masers provide estimates of the spiral arm velocities and their structure. I will discuss the techniques and properties of the MMB survey, before outlining recent results, which include the identification of regions of enhanced star formation believed to be indicative of the origins of the spiral arms and the interaction of the Galactic bar with the 3-kpc arms. I will also discuss the various follow-up programmes including a study of magnetic fields through associated hydroxyl masers.
\end{abstract}


\section{Introduction}
The Methanol Multibeam (MMB) survey was conceived (by R. J. Cohen and J. L. Caswell together with M. Sinclair, see the proceeding by J. L. Caswell for further details on the historical aspect of the project)  as the first unbiased, systematic Galactic plane survey for 6.7-GHz methanol masers. 6.7-GHz methanol masers are exclusively associated with high-mass star formation (Minier et al. 2003, Xu et al. 2008) and so are constrained to spiral arms and prominent features of Galactic structure. We expect high densities of high-mass star formation at the origins of the spiral arms and the ends of the bar and the central velocity of 6.7-GHz methanol maser emission is typically within 3 km/s of the systemic velocity (Szymczak et al., 2007, Pandian et al. 2009). These factors combine to make these masers ideal tracers of the structure and dynamics of the Milky Way, in addition to their benefits to understanding the processes of high-mass star formation.

\begin{figure}
\resizebox{\hsize}{!}{\includegraphics{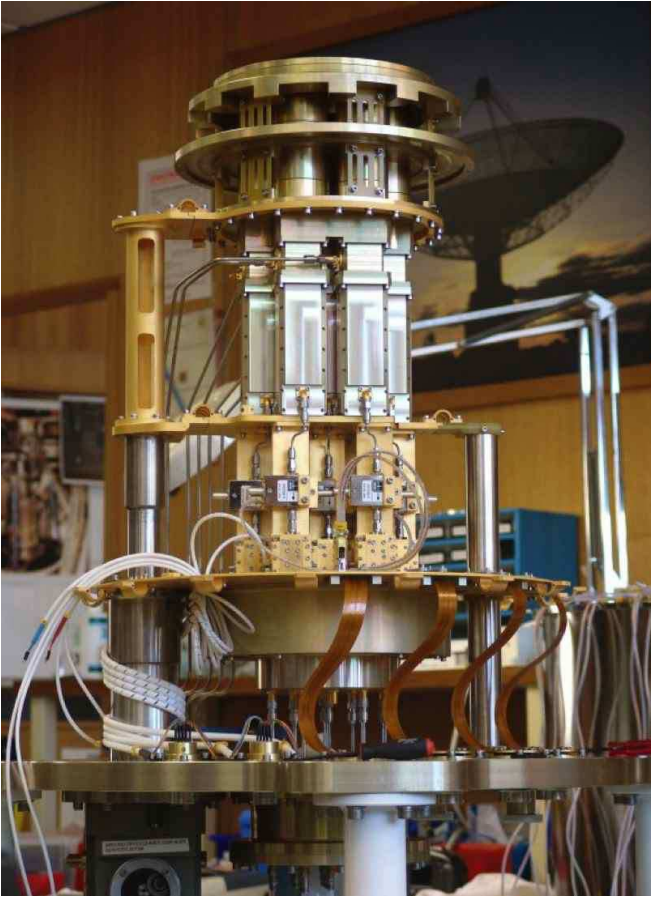}}
\caption{The purpose built seven beam Methanol Multibeam receiver.}
\end{figure}

\section{Observations}
The observations involved the initial blind survey with the Parkes Radio Telescope using the purpose-built 7-beam receiver shown in Figure 1 and covering a latitude range of $\pm$2$^{\circ}$. These were followed by high-resolution observations with the Australia Telescope Compact Array in order to position (to within 0.4 arcsec) any new sources or sources without previous high-resolution positions. Finally, all detections were re-observed with the Parkes Telescope again, at the precise positions, to obtain high signal--to--noise spectra (termed `MX' observations). The main survey observations were made between 2006 January and 2007 December, whilst the MX observations were made between 2007 January and 2009 April. The techniques of the MMB survey are detailed in full by Green et al. (2009a).

\begin{figure}
\centering{\includegraphics[width=80mm, clip]{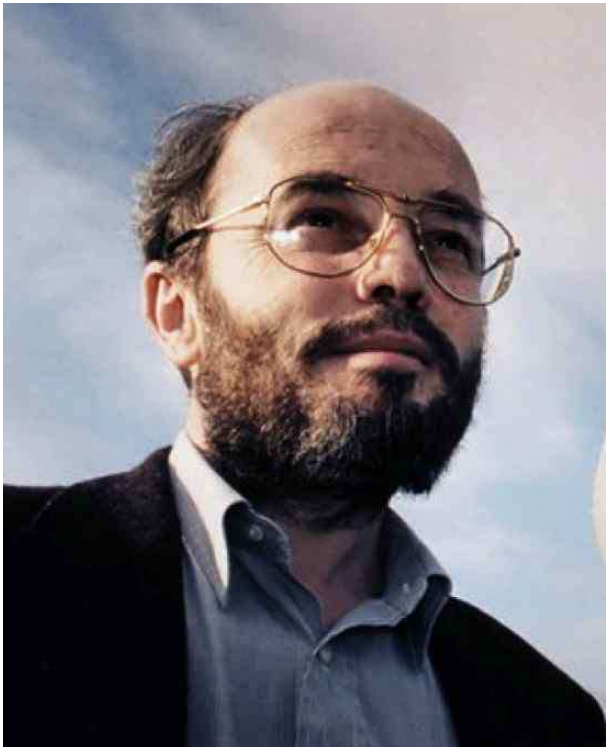}}
\caption{Jim Cohen 1948 -- 2006}
\end{figure}

\section{Survey Results}
The completion of the survey resulted in the detection of $\sim$1000 Galactic 6.7-GHz methanol masers and thus established an important legacy resource.
The survey catalogue is being released sequentially, with the first region covering Galactic longitudes 345$^{\circ}$ to 6$^{\circ}$(Caswell et al. 2010), the second Galactic longitudes 6$^{\circ}$ to 20$^{\circ}$ (Green et al. 2010), the third Galactic longitudes 330$^{\circ}$ to 345$^{\circ}$ (Caswell et al. 2011), and the fourth concludes the southern hemisphere portion of the catalogue, presenting detections within Galactic longitudes 186$^{\circ}$ and 330$^{\circ}$ (Green et al. 2012).  The last, more northerly region, Galactic longitudes 20$^{\circ}$ and 60$^{\circ}$, is currently in preparation. An online database has been set up at www.astromasers.org. 

Follow-up observations of the 12.2-GHz methanol maser transition have been made and Galactic longitudes 330$^{\circ}$ to 10$^{\circ}$ are presented in Breen et al. (2012) and 186$^{\circ}$ to 330$^{\circ}$ in Breen et al. (in prep.). 

\section{Survey Science}
One of the first scientific results of the survey was from the completion of a blind survey of the Small and Large Magellanic Clouds (observed during the non-Galactic hours between the main survey observations). A new methanol and excited-state hydroxyl maser were discovered (bringing the respective totals to four and two). Comparison of the abundances of masers within the Large Magellanic Cloud and the Milky Way shows maser populations are a factor of 10-50 times smaller in the Large Magellanic Cloud. In more detail, the hydroxyl and water maser populations are about 10 times smaller than our Galaxy and thus accounted for by the smaller star formation rate in the Large Magellanic Cloud. In contrast methanol is underabundant by a factor of 50, with the additional factor attributed to the lower metallicity of the Magellanic Clouds (Green et al. 2008).
 
The second result was a simple study of the 3-kpc arms, an inner structure of the Galaxy, located between the Galactic bars and the spiral arms. The near arm has long been known (van Woerden et al. 1957) but the far-side counterpart was only recently discovered (Dame \& Thaddeus 2008). There has been a continuing question over the existence of star formation within these structures, although some evidence (e.g. Caswell and Haynes, 1987) existed. However, both arms were shown by 6.7-GHz methanol maser detections to clearly exhibit high-mass star formation (Green et al. 2009b).

Following on from the previous result, 6.7-GHz methanol masers were shown to highlight regions of enhanced star formation indicating the starting points of the spiral arms and the interaction of the (long, thin) Galactic bar (Green et al. 2011). Furthermore masers also provide many candidates for tracing the full 3-kpc arm structure (a continuous ellipse rather than separate near and far components).

Additionally, the 12.2-GHz methanol companions detected in the follow-up observations were found to have smaller velocity ranges and lower flux densities than their 6.7-GHz methanol counterparts (Breen et al. 2011). There was an 80\% coincidence in velocity of the strongest maser features of the 6.7-GHz and 12.2-GHz transitions. The 12.2-GHz are associated with a later evolutionary stage and an increase in velocity range and luminosity with age was found for both transitions.

A comparison of the sites of 6.7-GHz methanol maser emission with infrared emission as detected by the GLIMPSE survey has been made, and found $>$80\% of the maser sites had associated infrared emission (Gallaway et al. in prep.).

\section{The Future}
The bright and compact nature of the maser sources makes them ideal for astrometry with Very Long Baseline Interferometry (VLBI). Such measurements provide accurate distances ($\pm$10 micro arcsec parallaxes) and full three-dimensional motion ($\pm$1 km/s). Maser astrometry can provide valuable data for both the structure of our Galaxy and its velocity field. The NRAO Very Long Baseline Array, European VLBI Network and Japanese VLBI Network are currently being used, with $\sim$30 parallaxes now published, and southern hemisphere observations are underway with the Long Baseline Array.

Another area that this project has provided the foundation for is in the study of magnetic fields. Although the 6.7-GHz methanol maser itself can not easily be used for measuring in situ magnetic fields, due to the molecule being non-paramagnetic, the often associated hydroxyl maser can be. As such all the detections of the MMB have been observed with the Australia Telescope Compact Array for the ground-state transitions of hydoxyl (as part of the `MAGMO' project) and measurements of the Zeeman splitting of components will provide estimates of the in situ magnetic field strength and line-of-sight orientation.

Finally it is the hope of the collaboration to follow up the survey with a northern hemisphere counterpart.\\

\acknowledgments
The MMB project is dedicated to Jim Cohen who sadly passed away in 2006.\\

We are grateful to all the staff and engineers at both the Parkes Radio Telescope and the Australia Telescope Compact Array.

{\it Facilities:} \facility{Parkes}, \facility{ATCA}.

\end{document}